\documentclass{sig-alternate-05-2015}
\usepackage{alltt                                    
          , multirow
          , booktabs
          , listings
          , graphicx
          ,float
	,cite
          ,verbatim
         ,mathtools
	,url
	,amsmath
}
\usepackage[table]{xcolor}
\usepackage[numbers]{natbib}     
\usepackage{syntax}
\usepackage{algorithm}
\usepackage{enumitem}
\usepackage{threeparttable}
\usepackage{pbox}
\usepackage{balance}
\usepackage{algpseudocode}
\usepackage{hhline}

\usepackage{expl3}
\ExplSyntaxOn
\newcommand\latinabbrev[1]{
  \peek_meaning:NTF . {
    #1\@}%
  { \peek_catcode:NTF a {
      #1., \@ }%
    {#1., \@}}}
\ExplSyntaxOff

\newcommand{\CASE}[1]{\STATE \textbf{case} #1\textbf{:} \begin{ALC@g}}
\newcommand{\ENDCASE}{\end{ALC@g}}

\newcommand{\DEFAULT}{\STATE \textbf{default:} \begin{ALC@g}}
\newcommand{\ENDDEFAULT}{\end{ALC@g}}
\newcommand{\DEFAULTLINE}[1]{\STATE \textbf{default:} }
\let\footnotesize\scriptsize

\newsavebox{\supbox}
\newcommand{\bsup}{\begin{lrbox}{\supbox}$\tt\scriptstyle}
\newcommand{\esup}{$\end{lrbox}{}^{\usebox{\supbox}}}
\def\eg{\latinabbrev{e.g}}
\def\ie{\latinabbrev{i.e}}

\algnewcommand{\LineComment}[1]{\State \(\triangleright\) #1}

\renewcommand{\refname}{REFERENCES}

\definecolor{lightpurple}{rgb}{0.8,0.8,1}
\definecolor{codebg}{RGB}{255,255,255}
\definecolor{commentcolor}{RGB}{11,140,11}
\begin{document}
\toappear{}
\setcopyright{acmcopyright}
%
\conferenceinfo{ASE}{'2016 Singapore}


\title{QUICKAR: Automatic Query Reformulation for Concept Location using Crowdsourced Knowledge }
%
%
%
%
%
%

\numberofauthors{2} 
%
\author{
%
%
\begin{tabular}[t]{@{}c@{}}
Mohammad Masudur Rahman~~~~Chanchal K. Roy\\
       \affaddr{Department of Computer Science, University of Saskatchewan, Canada}\\
       \email{\{masud.rahman, chanchal.roy\}@usask.ca}
\end{tabular}
}

\maketitle

\vspace{-1cm}

\begin{abstract}
During maintenance, software developers deal with numerous change requests made by the users of a software system.
Studies show that the developers find it challenging to select appropriate search terms from a change request during concept location.  
In this paper, we propose a novel technique--QUICKAR--that automatically suggests helpful reformulations for a given query by leveraging the crowdsourced knowledge from Stack Overflow.
It determines semantic similarity or relevance between any two terms by analyzing their adjacent word lists from the programming questions of Stack Overflow, and 
then suggests semantically relevant queries for concept location.
Experiments using 510 queries from two software systems suggest that our technique
can improve or preserve the quality of 76\% of the initial queries on average which is promising.
Comparison with one baseline technique validates our preliminary findings, and also demonstrates the potential of our technique.

\end{abstract}







\section{Introduction}\label{sec:introduction}
Studies show that about 85\%--90\% of the total effort is spent in software maintenance and evolution \cite{legacy,modernizing}.
During maintenance, software developers deal with numerous change requests made by the users of a software system.
Although the users might be familiar with the application domain of the software, they generally lack the idea of how a particular software feature is implemented in the source code.
Hence, the requests from them generally involve domain related concepts (\eg\ application features), and they are written in an unstructured fashion using natural language texts.
The developers need to prepare appropriate search query from those concepts, and then identify the relevant location(s) in the code to implement the requested change(s). Unfortunately, preparing such a query is highly challenging and error-prone for the developers \cite{kevic,vocaprob}. 
Based on a user study, \citet{kevic} report that developers were able to suggest good quality search terms for only 12.2\% of the change tasks.
\citet{vocaprob} suggest that there is a little chance (\ie\ 10\%--15\%) that developers guess the exact words used in the source code. 
One way to assist the developers in this regard is to automatically suggest helpful reformulations for the initially executed query.   

Existing studies use relevance feedback from developers \cite{relevancefb} or information retrieval techniques \cite{refoqus}, query quality \cite{specificity,soniaase} and the context of query terms within the source code \cite{ccmapping,infer} in suggesting reformulated queries.  
\citet{relevancefb} make use of explicit feedback on document relevance from the software developers, and then suggest reformulated queries using Rocchio's expansion.
\citeauthor{refoqus} and colleagues \cite{refoqus,qeffect,soniaase,soniatool,specificity} take quality of the query into consideration, and suggest the best reformulation strategy for a given query using machine learning.
\citet{ccmapping} analyze leading comments and method signatures from the source code for mining semantically similar word pairs, and then suggest reformulated query using those word pairs. 
While these above techniques are reported to be novel or effective, they are also limited in certain aspects.
First, collecting explicit feedback from the developers could be highly expensive, and such study \cite{relevancefb} could also be hard to replicate. 
Second, machine learning model of \citeauthor{refoqus} is reported to be performing well in the case of \emph{within-project} training, and only 51--72 queries are considered from each of the five projects for training and testing \cite{refoqus}. 
Given such small dataset, the reported performance possibly could not be generalized for large systems. 
Third, \citeauthor{ccmapping} require the source code to be well documented for the mining of word pairs, and hence, might not perform well if the code is poorly documented \cite{ccmapping}.
Thus, we need a technique that is neither subject to the training data nor the availability of comments in the source code.
One way to possibly overcome those concerns is to apply \emph{crowd generated knowledge} in the reformulation of queries for concept location.

In this paper, we propose a novel technique--QUICKAR--that automatically identifies semantically similar words to an initial query not only from project source code but also from crowdsourced content of Stack Overflow Q \& A site, and then suggests a reformulated query.
The technique collects adjacent word lists from the programming questions of Stack Overflow for any two terms, and determines their semantic similarity by comparing their corresponding adjacency lists \cite{wordsim}.
In short, QUICKAR not only follows the essence of a \emph{nearest neighbour classifier} \cite{nnc} in the context of natural language texts but also harnesses the technical corpus developed by a large crowd over the years in estimating \emph{semantic similarity} or \emph{relevance}. 
Such simple but intuitive estimation of semantic relationship could be highly useful for suggesting an alternative version of a given query. 
QUICKAR also addresses the overarching \emph{vocabulary mismatch problem} \cite{vocaprob}.
First, Stack Overflow is curated by a large crowd of four million technical users, and the millions of questions and answers posted by them are a great source for technical vocabulary (\eg\ API names) \cite{rigby}.
Thus, QUICKAR mines semantically similar words not only from a larger corpus (\ie\ compared to a single project \cite{infer}) but also from a more appropriate vocabulary (\ie\ compared to WordNet \cite{sridhara,infer}).
Second, \citet{rack} reported a significant overlap (\ie\ 73\%) between the vocabulary of real life code search queries and that of question titles from Stack Overflow.
QUICKAR mines 500K programming related questions from Stack Overflow carefully, and reaps the benefit through meaningful vocabulary extension.  
To the best of our knowledge, no existing studies apply crowdsourced knowledge yet in the reformulation of queries for concept location which makes our technique novel.

Experiments using 510 concept location queries from two subject systems--\texttt{ecf} and \texttt{eclipse.pde.ui}--suggest that our technique can improve or preserve the quality of 76\%  (\ie\ improves 66\% and preserves 10\%) of the initial queries through reformulation which is promising according to relevant literature \cite{refoqus,relevancefb}.
Comparison with one baseline technique--Rocchio's expansion \cite{refoqus,survey}--validates our preliminary findings, and also demonstrates the potential of our technique for query reformulation.
While the preliminary findings are promising, they must be validated using further experiments. In this paper, we make the following contributions:
\begin{itemize}[noitemsep,topsep=1pt]
\item Construction of a word adjacency list database by mining 500K questions from Stack Overflow for the estimation of semantic similarity or relevance between words.
\item A novel technique that suggests helpful reformulations for a given query for concept location by leveraging crowdsourced knowledge from Stack Overflow.   
\end{itemize}


\section{Motivating Example} \label{sec:motivation}
Software change requests and project source code are often written by different people, and they use different vocabularies to describe the same technical concept. Concept location community has termed it as \emph{vocabulary mismatch problem} \cite{vocaprob,refoqus}.
Table \ref{table:duplicate} shows three different questions from Stack Overflow that are marked as \emph{duplicates} or \emph{linked} by the users of the site.
These questions use three different verbs--\emph{`create', `cause'} and \emph{`track'}-- to describe the same programming issue--\emph{locating memory leaks}, and this can be considered as a real life parallel for vocabulary mismatch issue.   
Although, these verbs have different semantics in English literature, they share the same or almost similar semantic in this technical literature-- programming questions \cite{infer}. 
More interestingly, their semantic similarities can also be approximated from their adjacent word lists. 
In graph theory, two nodes are considered to be \emph{connected} if they share the same neighbour nodes \cite{nnc}. We adapt that idea for natural language texts, and apply to the estimation of semantic connectivity between words. 
The co-occurred word lists (\ie\ sentence as a context unit) of \emph{`create', `cause'} and \emph{`track'}--\{\emph{memory, leak, Java}\}, \{\emph{easiest, way, memory, leak, Java}\} and \{\emph{down, memory, leak, garbage, collection, issues, Java}\}--respectively
share multiple words among themselves. Thus, comparison between any two such lists can potentially approximate the semantic similarity or relevance of their corresponding words \cite{wordsim}. 
In this research, we apply the above methodology in semantic similarity estimation, and then use the similar terms for the reformulation of an initial query.

\begin{figure*}[!t]
\centering
\includegraphics[width=7in ]{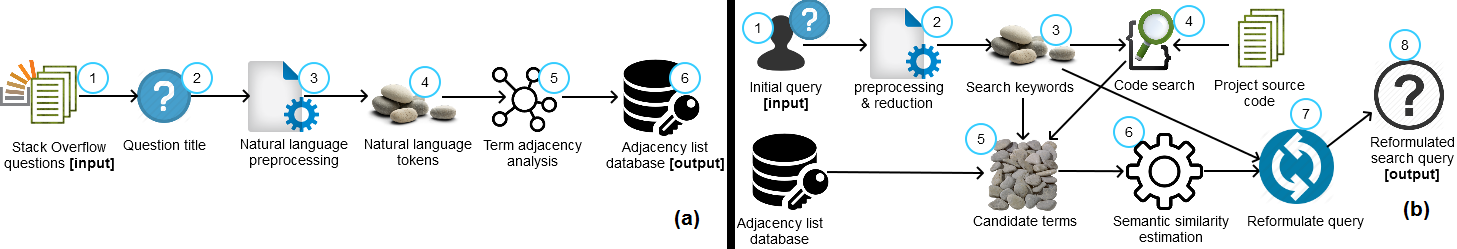}
\centering
\vspace{-.8cm}
\caption{Proposed technique for query reformulation--(a) Construction of word adjacency list database from Stack Overflow questions, and (b) Reformulation of an initial query for concept location}
\label{fig:sysdiag}
\vspace{-.5cm}
\end{figure*}

\begin{table}[!t]
\centering
\caption{Duplicate Questions from Stack Overflow}\label{table:duplicate}
\resizebox{3.35in}{!}{%
\begin{threeparttable}
\begin{tabular}{c|l}
\hline
\textbf{ID} & \textbf{Title of Question}\\
\hline
\hline
6470651 & Creating a memory leak with Java \\
\hline
4948529 & Easiest way to cause memory leak in Java?\\
\hline
1071631 & Tracking down a memory leak/garbage-collection issue in Java\\
\hline
\end{tabular}
\end{threeparttable}
}
\vspace{-.7cm}
\end{table}


\vspace{-.2cm}
\section{QUICKAR: Query Reformulation using Crowdsourced Knowledge}\label{sec:proposed}
Programming questions and answers from Stack Overflow were previously mined for API elements \cite{rigby,rack}, Q \& A dynamics \cite{nasehi,west} or post suggestions \cite{seahawk,surfclipse}. In this research, we make a novel use of such questions in query reformulation for concept location.
Since these questions contain unstructured information, relevant items (\ie\ potential alternative query term) should be carefully extracted and then applied to query reformulation.
We first construct a database containing adjacent word list for each of the individual words taken from the title of the questions, and then leverage that information in suggesting reformulated queries. 
Fig. \ref{fig:sysdiag} shows the schematic diagram of our proposed technique--QUICKAR--for query reformulation. 
Thus, our technique can be divided into two major parts--(a) Construction of an adjacency list database from Stack Overflow questions, and (b) Reformulation of an initial query--as follows:

\subsection{Construction of Adjacency List Database}\label{sec:database}
Word co-occurrence is often considered as a proxy for semantic relevance between words in natural language texts \cite{rada}. 
\citet{wordsim} first propose to use contextual words (\ie\ word co-occurrence) from the programming questions and answers of Stack Overflow in identifying semantically similar software-specific word pairs. 
While they introduce the idea, we adapt that idea for a specific software maintenance task--search query reformulation for concept location.
Given the significant overlap (\ie\ 73\%) between real life code search queries and titles of Stack Overflow questions \cite{rack}, we collect the titles of 500K Java related questions from Stack Overflow using Stack Exchange Data Explorer\footnote{http://data.stackexchange.com/stackoverflow}.
We check for \texttt{<java>} tag in the tag list for identifying the Java related questions. 
We perform standard natural language preprocessing (\ie\ word splitting, stop word removal), and turn each of those titles into a sequence of words (Step 3, 4, Fig. \ref{fig:sysdiag}-(a)).
We decompose each camel case word (\eg\ \texttt{GenericContainerInstantiator}) into separate tokens (\eg\ \texttt{Generic, Container, Instantiator}), and apply a standard list\footnote{https://code.google.com/p/stop-words/}  for stop word removal.
It should be noted that we avoid stemming to ensure a meaningful reformulation of the query.
After preprocessing step, we found a large set of 81,394 individual words (\ie\ 2,660,257 words in total) from our collected titles.
Given that source code of a software project is often authored by a small group of developers, such a large set could possibly \emph{extend} the vocabulary of the code. 
We then use a \emph{sliding window} of \texttt{window size = 2} to capture co-occurred words from the titles \cite{rada}, and construct an adjacency list for each of the individual words. 
For example, based on Table \ref{table:duplicate}, the adjacency list for the word \emph{`memory'} would be--\{\emph{creating, leak, cause, down}\}.  
We collect adjacency list for each of the 81,394 words where each list contains 21 co-occurred words on average (Step 5, 6, Fig. \ref{fig:sysdiag}-(a)).
These lists comprise our database which is later accessed frequently for query reformulation.

\subsection{Reformulation of an Initial Query}\label{sec:reformulation}
Fig. \ref{fig:sysdiag}-(b) shows the schematic diagram and Algorithm \ref{algo} shows the pseudo code for our query reformulation technique--QUICKAR.
We collect semantically similar words (\ie\ with initial query) from two different but relevant sources--project source code and Stack Overflow questions, and then
reformulate a given query for concept location. We discuss different intermediate steps involved in such reformulation as follows:

\textbf{Collection of Candidate Terms:} Existing literature mostly relies on the source code of a software system \cite{refoqus,ccmapping,infer} or WordNet \cite{sridhara} for reformulated query suggestion.
Unfortunately, source code often might not contain a rich vocabulary \cite{ccmapping} and WordNet might also not be appropriate for technical words \cite{sridhara,wordsim}. 
We thus collect candidate terms for possible query expansion not only from the source code but also from another technical literature--programming questions of Stack Overflow (Line 6--Line 8, Algorithm \ref{algo}). 
In the case of project source, we perform code search using the reduced keywords ($K$) from the initial query ($Q$).
We use \emph{Apache Lucene} \cite{seahawk}, a popular implementation of Vector Space Model (VSM), for code search, and then collect the Top-5 (\ie\ cut off) retrieved documents as the source for candidate terms.
The cut-off value is chosen based on iterative experiments.
We perform standard natural language preprocessing on those documents, and extract each of the terms as the reformulation candidates ($T_p$) (Step 3--5, Fig. \ref{fig:sysdiag}-(b)).
Relevant literature \cite{refoqus} also follows the same procedure for candidate term selection.
In the case of questions from Stack Overflow, we collect such words as candidates ($T_{so}$) that frequently co-occurred in those questions with the keywords from the initial query. 
The underlying idea is that if two words frequently co-occur in various technical contexts, they share their semantics and thus, are possibly semantically relevant \cite{rada,wordsim}. 
We use our adjacency list database (Section \ref{sec:database}) for identifying the second set of candidates.

\begin{algorithm}[!t]
\caption{Query Reformulation using Crowd Knowledge}
\label{algo}
\begin{algorithmic}[1]
\Procedure{QUICKAR}{$Q$}\Comment{$Q$: initial search query}
\State $Q' \gets$ \{\}\Comment{reformulated search query}
\LineComment{collecting keywords from the initial search query}
\State $K \gets$ collectKeywords($Q$)
\State $K \gets$ reduceKeywords($K$)
\LineComment{collecting candidate terms}
\State $T_p \gets$ getCandidateTermsFromProject($K$)
\State $T_{so} \gets$ getCandidateTermsFromSO($K$)
\LineComment{estimating semantic similarity of the candidates}
\For{Candidate $T_i$ $\in$ $T_p$}
\State $Adj_{T_i} \gets$ getAdjacencyListFromDB($T_i$)
\For{Keyword $K_j \in K$}
\State $Adj_{K_j} \gets$ getAdjacencyListFromDB($K_j$)
\LineComment{contextual similarity between words}
\State $S_{cos} \gets$ getCosineSimilarity($Adj_{T_i}, Adj_{K_j}$) 
\State $R_p[T_i].score \gets R_p[T_i].score + S_{cos}$
\EndFor
\EndFor
\For{Candidate $T_i$ $\in$ $T_{so}$}
\For{Keyword $K_j \in K$}
\LineComment{co-occurrence between words}
\State $S_{cof} \gets$ getCo-occurrenceFreq($T_i, K_j $) 
\State $R_{so}[T_i].score \gets R_{so}[T_i].score + S_{cof}$
\EndFor
\EndFor
\LineComment{ranking and selection of candidates}
\State $SR_{Top_p} \gets$ selectTopK(sortByScore($R_p$))
\State $SR_{Top_{so}} \gets$ selectTopK(sortByScore($R_{so}$))
\State $R \gets$ selectiveCombine($SR_{Top_p}, SR_{Top_{so}}$) 
\LineComment{reformulate the initial query}
\State $Q' \gets$ selectiveReformulate($Q, K, R$)
\State \textbf{return} $Q'$
\EndProcedure
\end{algorithmic}
\end{algorithm}
\setlength{\textfloatsep}{2pt}

\textbf{Estimation of Semantic Similarity or Relevance:} Since we target to reformulate a query using meaningful alternatives, we need to choose such terms from the candidates that are either semantically similar or highly relevant to the initial query.
\citeauthor{wordsim} use contextual words (\ie\ based on co-occurrence) from Stack Overflow for automatically identifying similar software-specific words.
In the context of query reformulation for software maintenance, we similarly apply adjacency list to the estimation of semantic relevance between two terms.
We collect adjacency list (\ie\ from the adjacency list database) for each of the candidate terms and the keywords of the initial query, and estimate their similarities using \emph{cosine similarity} measure ($S_{cos}$).
Cosine similarity is frequently used in information retrieval for determining textual similarity between two given documents. It returns a value between zero (\ie\ completely dissimilar) and one (\ie\ completely similar).
Thus, a candidate term achieves score only if it shares its context (\ie\ adjacent word list, $Adj_{T_i}$) with that (\ie\ $Adj_{K_j}$) of the keywords across various questions from Stack Overflow.  
QUICKAR iterates this process for each of the candidates ($T_p$), and accumulates their similarity scores against the keywords (Line 9--Line 18, Algorithm \ref{algo}, Step 6, Fig. \ref{fig:sysdiag}-(b)).
We also determine co-occurrence frequency ($S_{cof}$) between each candidate term and each keyword in the titles of Stack Overflow questions \cite{rack}, and derive another set of scores for the candidates ($T_{so}$) (Line 19--Line 25, Algorithm \ref{algo}, Step 6, Fig. \ref{fig:sysdiag}-(b)).
Finally, we end up with two sets of candidates (\ie\ collected from two different sources), and QUICKAR determined their relevance to the initial query in terms of their context or direct co-occurrences in the Stack Overflow questions.  

\begin{table}[!t]
\centering
\caption{Experimental Dataset }\label{table:dataset}
\resizebox{3.35in}{!}{%
\begin{threeparttable}
\begin{tabular}{l|c|c|c|c}
\hline
\textbf{Subject System} & \textbf{Release ID} & \textbf{\#Files} & \textbf{\#Methods} & \textbf{\#Queries}\\
\hline
\hline
\texttt{ecf} & 170_170 & 5,781& 21,447 & 222  \\
\hline
\texttt{eclipse.pde.ui} & I20151110-0800 & 7,579& 31,468  & 288 \\
\hline
\textbf{Total} & -- & 13,360 & 52,915 & 510 \\

\hline
\end{tabular}
\centering
\end{threeparttable}
}
\end{table}

\textbf{Candidate Term Ranking \& Top-K Selection :} Once relevance estimates for both candidate sets--$R_p$ and $R_{so}$--are collected, they are ranked based on their estimates.
We then collect the Top-$K=5$ candidates from each ranked list, selectively choose the top candidates ($R$) from both lists, and then treat them as similar or relevant to the initial query ($Q$) (Line 26--Line 29, Algorithm \ref{algo}).  
In particular, the \emph{nominal terms} (\ie\ nouns) from both lists are chosen \cite{rack}.
Thus, we select such terms for reformulation that co-occur with the initial query keywords not only in the project source but also in the titles of the questions from Stack Overflow.  

\textbf{Query Reduction \& Expansion:} We apply both reformulation strategies--\emph{reduction} and \emph{expansion}--to the initial query \cite{survey}.
In the case of \emph{reduction}, we apply a conservative strategy as was also applied by \citeauthor{refoqus}. 
We discard the keywords from initial query that either are \emph{non-nominal} \cite{rack} or occurred in more than 25\% of the documents of the project corpus. 
Such keywords are not specific enough and thus, are not useful for document retrieval \cite{refoqus}.
In the case of \emph{expansion}, we apply the semantically similar or relevant terms ($R$) returned by QUICKAR to the query. If there exist $M$ terms after the reduction step (\ie\ Line 5), we append $(10-M)$ relevant terms from $R$ to the query, and prepare an alternative query ($Q'$)
(Line 29--Line 32, Algorithm \ref{algo}, Step 7, 8, Fig. \ref{fig:sysdiag}-(b)).
We also decompose each camel case term into separate tokens, and preserve both the separate and the camel case terms into the reformulated query \cite{refoqus}.
It should be noted that if our reduction step already improves the initial query, we conveniently avoid its expansion. 
 
\textbf{Working Example:} Let us consider a change request (ID: 408030) from \texttt{ecf} project.
We select title of the request--\emph{``RestClientService ignores content encoding"}--as the initial search query for the request, as was also used by the existing literature \cite{refoqus}. 
The query returns the first relevant document (\ie\ Java class) at 44$^{th}$ position when tested using \emph{Apache Lucene}. 
On the other hand, QUICKAR returns the reformulated query--\emph{``Rest Client Service RestClientService content Web Java Executor WebService Http"}--that returns the same relevant document at 1$^{st}$ position in the search result.
Please note that our technique expands the initial query using several relevant terms such as \emph{``WebService", ``Executor"} and \emph{``Http"} and also discards some terms such as ``\emph{encoding}" or ``\emph{ignore}", which improved the query. 
We also expand the query simply using terms from project source--\emph{``RestClientService ignores content encoding call container service http test"}--that returns the document at 11$^{th}$ position.
This clearly demonstrates that candidates from the source code of a project might always not be sufficient, and crowd generated vocabulary from Stack Overflow can complement them, and thus, can assist in effective query reformulation.

\begin{table*}[!t]
\centering
\caption{Performance of QUICKAR}\label{table:result}
\resizebox{7in}{!}{%
\begin{threeparttable}
\begin{tabular}{l|c|c|c|c|c|c|c|c||c|c|c|c|c|c|c||c}
\hline
\multirow{2}{*}{\textbf{System}} & \multirow{2}{*}{\textbf{\#Queries}} &\multicolumn{7}{c||}{\textbf{Improvement}} & \multicolumn{7}{c||}{\textbf{Worsening}} & \textbf{Preserving}   \\
\hhline{~~---------------}
& & \#Improved & Mean & Q1 & Q2 & Q3 & Min. & Max. & \#Worsened & Mean & Q1 & Q2 & Q3 & Min. & Max. & \#Preserved\\
\hline
\hline
\textbf{ecf} & 222 & \textbf{159 (71.62\%)} & 194 & \textbf{10} & 27 &  156 & 1 & 2335 & 41 (18.47\%) & 764 & 67 & 245 & 1173 &  18 & 4330 & \textbf{22 (9.90\%)} \\
\hline
\textbf{pde.ui}  & 288 & \textbf{177 (61.46\%)} & 219 & \textbf{17} &  51 & 171 & 1 & 4766 & 83 (28.82\%) & 558 & 93 & 244 & 630 & 17 & 2996 & \textbf{28 (9.72\%)}\\
\hline
& Total=510 & \textbf{Avg=66.54\%}  & &  &  &  &  &  & Avg=23.65\%  &  &  &  &  &  &  & \textbf{Avg=9.81\%}    \\
\hline
\end{tabular}
\centering
\textbf{pde.ui}=\texttt{eclipse.pde.ui}, \textbf{Mean}=Mean rank of first relevant document in the search result, \textbf{Q$_i$}= Rank value for $i^{th}$ quartile of all result ranks
\end{threeparttable}
}
\vspace{-.7cm}
\end{table*}

\vspace{-.2cm}
\section{Experiment}\label{sec:experiment}
One of the most effective ways to evaluate a query reformulation technique is to check whether the reformulated query improves the search results or not. 
We define improvement of search results as the bubbling up of the first relevant document to the top positions of the result list \cite{refoqus}.  
That is, a good reformulation of query provides a better rank than a baseline query for the first relevant document.   
We conduct experiments using 510 change requests from two Java subject systems of \emph{Eclipse}--\texttt{ecf} and \texttt{eclipse.pde.ui}-- and a well known search engine--\emph{Apache Lucene} \cite{seahawk,refoqus}.
We also compare with a baseline technique for query reformulation to validate our findings. In particular, we attempt to answer the following research questions using our experiments:
\begin{itemize}[noitemsep,topsep=1pt]
\item \textbf{RQ$\mathbf{_1}$:} How does QUICKAR perform in the reformulation of a query for concept location?  
\item \textbf{RQ$\mathbf{_2}$:} Can crowdsourced knowledge from Stack Overflow improve a given query significantly?
\item \textbf{RQ$\mathbf{_3}$:} How does QUICKAR perform compared to the baseline technique for query reformulation? 
\end{itemize}

\subsection{Experimental Dataset \& Corpus}

\textbf{Dataset Collection:} 
Table \ref{table:dataset} shows details of our selected subject systems.
We first collect the \emph{RESOLVED} change requests (\ie\ bug reports) from \emph{BugZilla} for each of the selected systems. Then we identify such commits that implemented those requests in their corresponding \emph{GitHub} repositories. 
We consider a commit as \emph{eligible} only if its title contains a specific request identifier (\eg\ Bug: 408030). 
This practice is common for relevant literature for evaluation \cite{reenact}.
The step provides 495 and 542 requests from \texttt{ecf} and \texttt{eclipse.pde.ui} respectively.
We also collect \emph{change set} from each of the identified commits which are later used as the \emph{solutions} for the corresponding change requests.
We then consider title of each request as the baseline query, and identify such queries that return their first relevant results with poor rank (\ie\ $rank$>10).
That is, the baseline query needs reformulation for returning a better rank. 
This filtration step left us with 222 and 288 baseline queries from \texttt{ecf} and \texttt{eclipse.pde.ui} respectively for the experiments.    


\textbf{Corpus Preparation:} Unlike an unstructured natural language document, a source code document contains items beyond regular texts such as \emph{classes, interfaces, methods} and \emph{constructors}.
One should consider such structures for effective retrieval of the source documents from a project.
We thus decompose each Java document into methods, and consider each of those methods as a single document of the corpus.
This step provides 21,447 and 31,468 Java methods from \texttt{ecf} and \texttt{eclipse.pde.ui} respectively.
We collect these methods using Javaparser\footnote{https://github.com/javaparser/javaparser}, and apply natural language preprocessing to them.
In particular, we remove all punctuation marks, Java programming keywords and English stop words from the body and signature of the method, and also decompose each camel case token into individual tokens.

\subsection{Evaluation of QUICKAR}
We execute each of our reformulated queries and baseline queries from each subject system with \emph{Apache Lucene}, and compare their topmost ranks for evaluation.
We identify the queries that were improved, worsened or preserved based on those ranks. Table \ref{table:result} reports the outcome of our preliminary investigation.
On average, QUICKAR was able to improve 66\% of the baseline queries while preserving the quality of 10\% which are highly promising according to relevant literature \cite{refoqus,relevancefb,ccmapping}.
We see that QUICKAR can return the top results within the 10$^{th}$ position for 25\% of 159 requests from \texttt{ecf} system. 
It also performs similarly for \texttt{eclipse.pde.ui}, and returns the top results within the 17$^{th}$ position. 
One might argue about the verbatim use of the title of a change request as the baseline query. 
However, we also experimented with preprocessed version (\ie\ stop word removal, camel case decomposition) of the title. 
We found that the preprocessing step discarded important information, and did not provide much improvement in the query quality, which possibly justifies our choice about baseline query. 
Thus, to answer \textbf{RQ$_1$}, our proposed technique for query reformulation--QUICKAR--improves or preserves 76\% of the 510 baseline queries which is promising.

\begin{table}[!t]
\centering
\caption{Role of Crowdsourced Knowledge from SO}\label{table:role}
\resizebox{3.35in}{!}{%
\begin{threeparttable}
\begin{tabular}{l|c|c|c}
\hline
\textbf{Technique} & \textbf{Improved} & \textbf{Worsened} & \textbf{Preserved}\\
\hline
\hline
 Baseline query (preprocessed) & 17.84\% & 9.90\% & \textbf{72.27}\%\\
\hline
QUICKAR$_P$ & 49.15\% & 48.41\% & 2.44\% \\
\hline
QUICKAR$_{SO}$ & 47.83\% & 49.91\% & 2.27\%  \\
\hline
QUICKAR$_{red} $ & 55.55\% & 24.46\% & 19.99\%  \\
\hline
QUICKAR$_{ALL}$ & \textbf{66.54}\% & 23.65\% & \textbf{9.81}\%\\
\hline
\end{tabular}
\centering
\end{threeparttable}
}
\end{table}

We investigate how different reformulation decisions influence the end performance of our technique, and Table \ref{table:role} reports our findings.
We first experimented using a preprocessed version of the baseline queries, and found that the preprocessing step did not improve the queries much (\ie\ only 18\% improvement).
Since candidate terms for reformulation are extracted from both project source code and Stack Overflow questions, we need to examine their impact in the reformulated queries.
When we rely solely on source code, QUICKAR$_P$ can improve 49\% of the queries but degrades the quality of 48\%. 
Although the candidate terms from Stack Overflow questions alone might not be sufficient (\ie\ QUICKAR$_{SO}$  improves 48\% and degrades 50\%), they definitely can complement the candidate terms from the project source which leads to overall query quality improvement (\ie\ QUICKAR$_{ALL}$ improves 66\%).
We also performed \emph{Mann Whitney U (MWU)}-test on the provided ranks by QUICKAR$_P$ and QUICKAR$_{ALL}$, and found that QUICKAR$_{ALL}$ returns the results at significantly higher ranks (\ie\ \emph{p-values} 0.007<0.05 and 0.001<0.05 for \texttt{ecf} and \texttt{eclipse.pde.ui} respectively, Table \ref{table:test}) in the result list.
The negative \emph{mean rank difference} (MRD) in Table \ref{table:test} suggests that QUICKAR$_{ALL}$ returns the results at relatively closer to the top of the list than the counterpart, \ie\ \emph{extent of result rank improvement}.  
Thus, to answer \textbf{RQ$_2$}, crowdsourced knowledge from Stack Overflow questions can significantly improve the quality of a baseline query during reformulation.

Since QUICKAR involves two steps during query reformulation -- \emph{reduction} and \emph{expansion}, an investigation is warranted on how these two steps impact the end performance. 
According to our preliminary investigation, the reduction step (\ie\ QUICKAR$_{red}$) dominates over expansion step especially with \texttt{ecf} system. One possible explanation could be that we used a smaller version of the adjacency list database constructed from 50,000 (\ie\ 10\%) questions from the dataset.
Since accessing a large database is time-consuming, we made this feasible choice during experiment. 
However, further investigation and experiments are essential to mitigate such concern which we consider as a future work.
In short, the potential of crowdsourced knowledge is yet to be explored.

\begin{table}[!t]
\centering
\caption{Comparison with Baseline Technique}\label{table:compare}
\resizebox{3.35in}{!}{%
\begin{threeparttable}
\begin{tabular}{l|l|c|c|c}
\hline
\textbf{Technique} & \textbf{System} & \textbf{Improved} & \textbf{Worsened} & \textbf{Preserved}\\
\hline
\hline
Rocchio's  & \texttt{ecf}  & 39.64\% & 59.46\% & <1.00\%\\
\hhline{~----}
Expansion \cite{survey} & \texttt{pde.ui}  & 40.63\% & 59.38\% & 0.00\% \\
\hline
\multirow{2}{*}{QUICKAR$_{P}$} &  \texttt{ecf} & 53.15\% & 43.69\% & 3.15\%  \\
\hhline{~----}
 & \texttt{pde.ui} & 45.14\% & 53.13\% & 1.74\%\\
\hline
\multirow{2}{*}{QUICKAR$_{ALL}$} &  \texttt{ecf} & \textbf{71.62}\% & 18.47\% & \textbf{9.90}\%  \\
\hhline{~----}
 & \texttt{pde.ui} &\textbf{61.46}\% & 28.82\% & \textbf{9.72}\%\\
\hline
\end{tabular}
\centering
\textbf{pde.ui}=\texttt{eclipse.pde.ui}
\end{threeparttable}
}
\vspace{-.6cm}
\end{table}

\begin{table}[!t]
\centering
\caption{Result of Mann Whitney U-Tests}\label{table:test}
\resizebox{3.35in}{!}{%
\begin{threeparttable}
\begin{tabular}{l|c|c||c|c}
\hline
\multirow{2}{*}{\textbf{Technique pair}}  & \multicolumn{2}{c||}{\textbf{ecf}} & \multicolumn{2}{c}{\textbf{eclipse.pde.ui}} \\
\hhline{~----}
 & \textbf{p-value} & \textbf{MRD} & \textbf{p-value} & \textbf{MRD}\\
\hline
\hline
QUICKAR$_{ALL}$ vs. QUICKAR$_{P}$  & \textbf{0.007}<0.05 & -248 & \textbf{<0.001} & -369   \\
\hline
QUICKAR$_{ALL}$ vs. QUICKAR$_{SO}$ &  0.115>0.05 & -109  & \textbf{<0.001} & -253\\
\hline
QUICKAR$_{ALL}$ vs. Rocchio \cite{survey} & \textbf{<0.001}  & -388   & \textbf{<0.001} & -332  \\
\hline
\end{tabular}
\centering
\textbf{MRD}= \textbf{M}ean \textbf{R}ank \textbf{D}ifference
\end{threeparttable}
}
\end{table}

\subsection{Comparison with Baseline Technique}\label{sec:compare}
Although the conducted evaluation demonstrates the potential of our proposed technique--QUICKAR, we still investigate to at least partially validate our performance. 
We compare with a baseline technique--Rocchio's expansion \cite{survey,refoqus}--that is reported to be effective for query reformulation.
Rocchio's method first collects candidate terms from the Top-K (\ie\ $K=5$) source code documents returned by a baseline query. Then, it selects the most important candidate terms ($t$) for reformulation by calculating their \emph{TF-IDF} in each of those Top-K documents ($D$) as follows: 
\begin{equation*}
\setlength{\abovedisplayskip}{0em}
\setlength{\belowdisplayskip}{0em}
Rocchio(t)=\sum_{d\in R} TFIDF(t,d)
\end{equation*}
We implemented Rocchio's method in our working environment, experimented on the same corpus and applied similar natural language preprocessing.    
Table \ref{table:compare} reports the comparative analysis between our technique and Rocchio's method.
We see that our technique can improve 60\%--70\% of the baseline queries from each of the subject systems whereas such measure for Rocchio's method is close to 40\%.
More importantly, QUICKAR degrades less number of queries compared to its counterpart. While Rocchio's method worsened the quality of 60\% of the baseline queries during reformulation, such measure for QUICKAR is between 18\% to 29\%.  
We also performed \emph{Mann Whitney U}-test on the returned result ranks from both techniques, and found that QUICKAR provides significantly better ranks (\ie\ \emph{p-values} <0.001 and <0.001 for \texttt{ecf} and \texttt{eclipse.pde.ui} respectively, Table \ref{table:test}) than Rocchio's expansion.  
We also compared using a equivalent variant of our technique--QUICKAR$_{P}$, and found that the variant still performed better than Rocchio's method for both subject systems.
All these above preliminary findings clearly demonstrate the potential of our proposed technique. Thus, to answer \textbf{RQ$_3$}, our technique--QUICKAR--performs significantly better than the baseline technique \cite{survey} in query reformulation for concept location.

\section{Related Work}\label{sec:related}
Existing studies from the literature use relevance feedback from developers \cite{relevancefb} or information retrieval techniques \cite{refoqus}, query quality \cite{specificity,soniaase} or the context of a query in the source code \cite{ccmapping,infer} for suggesting query reformulations.  
\citet{relevancefb} capture explicit feedback on document relevance from the software developers, and then suggest reformulated queries using Rocchio's expansion.
Although their adopted methodology is meaningful, capturing feedback from the developers could be expensive, and such study is often difficult to replicate.  
\citeauthor{refoqus} and colleagues \cite{refoqus,qeffect,soniaase,soniatool,specificity} analyze quality of the query, and suggest the best reformulation strategy for any given query using machine learning.
Although their reported performance is significantly higher, such performance 
might not be generalized for large systems given their use of small dataset (\ie\ only 51--72 queries from each system).
\citet{ccmapping} analyze leading comments and method signatures from the source code, and suggest reformulated queries by extracting semantically similar word pairs.
However, their technique requires the source code to be well documented, and thus, might not perform well with poorly documented code. 
On the other hand, our technique--QUICKAR--complements the source code vocabulary by capturing appropriate candidate terms from the programming questions of Stack Overflow. 
\citet{survey} conduct a survey on the automatic query expansion (AQE) mechanisms applied to information retrieval.
Rocchio's expansion is one of such mechanisms which was adapted by earlier studies \cite{relevancefb,refoqus} in the context of software engineering. 
We consider this mechanism as the baseline reformulation technique, and compared with it using experiments (Section \ref{sec:compare}).
\citet{wordsim} first apply contextual words from Stack Overflow questions and answers for identifying semantically similar software-specific words. 
While they introduce the idea, we successfully adapt that idea for a software maintenance task, \ie\ query reformulation for concept location.
From a technical point of view, we collect candidate query terms opportunistically not only from project source code but also from questions of Stack Overflow, and determine their relevance to the initial query using crowdsourced knowledge (\ie\ adjacency list database).
We then apply the most relevant terms from both source code and Stack Overflow to query reformulation, and such methodology was not applied yet by any existing studies.




\section{Conclusion and Future Work}\label{sec:conclusion}
Studies show that software developers face difficulties in preparing an appropriate search query from a change request during concept location.  
In this paper, we propose a novel technique--QUICKAR--that automatically suggests effective reformulations for an initial query by leveraging the crowd generated knowledge from Stack Overflow.
The technique collects candidate query terms from both project source and questions of Stack Overflow, and then determines their applicability for the reformulated query by applying the crowdsourced knowledge.
Experiments using 510 change requests from two software systems suggest that our technique
can improve or preserve the quality of 76\% of the baseline queries which is promising.
Comparison with one baseline technique also validates our preliminary findings.
While the preliminary findings are promising, further experiments and investigations are warranted.

\bibliographystyle{plainnat}
\setlength{\bibsep}{0pt plus 0.3ex}
\scriptsize
\bibliography{sigproc}  
%
%
\end{document}